\RequirePackage{fix-cm}

\documentclass[twocolumn,epjc3]{svjour3m}          

\RequirePackage[T1]{fontenc}

\smartqed  

\RequirePackage{graphicx}
\RequirePackage{mathptmx}      
\RequirePackage{flushend}
\RequirePackage[numbers,sort&compress]{natbib}
\RequirePackage[colorlinks,citecolor=blue,urlcolor=blue,linkcolor=blue]{hyperref}
\usepackage{amsmath}
\usepackage{amsfonts}
\usepackage{amssymb}
\usepackage{epstopdf}
\usepackage{slashed}
\usepackage{float}

\journalname{\textsl{DESY 17-183}}
\def\II{\hbox{{1}\kern-.25em\hbox{l}}}
\begin{document}

\title{Correction exponents in the Gross~-~Neveu~-~Yukawa model at~$1/N^2$.}


\author{Alexander N. Manashov\thanksref{e1,addr1,addr2}
        \and
        Matthias Strohmaier  \thanksref{e2,addr2} 
}

\thankstext{e1}{e-mail: alexander.manashov@desy.de}
\thankstext{e2}{e-mail: matthias.strohmaier@ur.de}

\institute{Institut f\"ur Theoretische Physik, Universit\"at Hamburg, Luruper Chaussee 149, D-22761 Hamburg,
Germany\label{addr1}
          \and
          Institut f\"ur Theoretische Physik, Universit\"at  Regensburg, D-93040 Regensburg, Germany\label{addr2}
}

\date{Received: date / Accepted: date}

\maketitle

\begin{abstract}
We calculate the critical exponents $\omega_\pm$ in the $d$-dimensional  Gross-Neveu model in $1/N$ expansion with
$1/N^2$ accuracy. These exponents are related to the slopes of the $\beta$-functions at the critical point in the
Gross~-~Neveu~-~Yukawa model. They have been computed recently to four loops accuracy. We checked that our results are in
complete agreement with the results of the perturbative calculations.
\end{abstract}

\section{Introduction}\label{sect:intro}

The Gross~-~Neveu (GN) model,  originally introduced in~\cite{Gross:1974jv} in 1974 as a toy model for a study of
dynamical symmetry breaking,  found its application in many areas of physics. The model describes a system of
fermion fields with a quartic interaction. It is renormalizable in two dimensions, asymptotically free and, moreover, admits
an exact solution~\cite{Zamolodchikov:1978xm}. Above two dimensions the $N$-component GN~-~model is renormalizable within the
$1/N$ expansion technique. It possesses a nontrivial  Wilson~-~Fisher
fixed point and gives an example of a conformal field theory (CFT) in $d$-dimensions. Basic critical indices of the GN model
are available at
$1/N^2$ order~\cite{Gracey:1990wi,Gracey:1992cp,Derkachov:1993uw,Vasiliev:1993pi,Gracey:1993kb}
 and the index $\eta$~--~the
anomalous dimension of the fermion field~--~at $1/N^3$~\cite{Vasiliev:1992wr,Gracey:1993kc}. These results were obtained by
methods of the self-consistency equation  and conformal bootstrap  developed
in~\cite{Vasiliev:1981yc,Vasiliev:1981dg,Vasiliev:1982dc}.
A recent revival of interest to this model is due to a relevance of fermionic systems with quartic interactions to the
description of the phase transition in graphene \cite{Janssen:2016xvc}.

The UV completion  of the GN model contains an additional scalar field  with  a quartic self-interaction and is known as the
Gross-Neveu-Yukawa (GNY) model \cite{ZinnJustin:1991yn}.  Moreover, the chiral extension of the GNY model -- the
Nambu-Jona-Lasinio (NJL) model -- describes, for a low number of fermion flavors,   a system with an emergent
supersymmetry~\cite{Fei:2016sgs}.
The recent calculation of the corresponding  renormalization group (RG) functions
with four-loop accuracy ~\cite{Karkkainen:1993ef,Mihaila:2017ble,Zerf:2017zqi}  confirms the emergence of supersymmetry at the
critical point.

The four--loop calculations are quite involved and carried out with the help of computer algebra. The results available from
$1/N$ expansion provide an additional check for the perturbative calculations. The  four loop RG
functions obtained in~\cite{Zerf:2017zqi} are in a perfect agreement with the results of $1/N$
calculations~\cite{Gracey:1990wi,Gracey:1992cp,Derkachov:1993uw,Vasiliev:1993pi,Gracey:1993kb,Vasiliev:1992wr,Gracey:1993kc}.
In this paper we present $1/N^2$ expressions for the other two indices -- the so-called correction exponents that are related to
the slopes of
$\beta$ functions at the critical point and were known only with $1/N$ accuracy ~\cite{Gracey:2017fzu}.

The $1/N$ calculations are usually done with the help of the methods of the self-consistency equations or conformal bootstrap
\cite{Vasiliev:1981yc,Vasiliev:1981dg,Vasiliev:1982dc}. These methods are very effective for the calculation
of the critical indices of the basic and auxiliary fields, but are not very suitable in the case of local operators.
Therefore we use a different method developed
in~\cite{Vasiliev:1975mq,Vasiliev:1993ux,Derkachov:1997ch}. A detailed description of the method can be found
in~\cite{Derkachov:1997ch,Derkachov:1998js} and an application to the GN model in~\cite{Manashov:2016uam}.

The paper is organized as follows: In sect.~\ref{GNY} we recall the formulation of the GNY model and show that the slopes of
$\beta$ functions at the critical point coincide with the critical dimensions of  certain operators of dimension four.
In sect.~\ref{largeN} we review briefly the  $1/N$ expansion technique for the GN model and present the rules for calculating
the anomalous dimensions of local operators. Section~\ref{diagrams} contains the details of the calculation of the correction
exponents at order~$1/N^2$.  In \ref{app:1/n^2indices} we  collect expressions for relevant basic indices. \ref{app:SEV}
contains details of the calculation of the self-energy and vertex correction diagrams and in \ref{app:part_results} we collect
results for the individual diagrams.

\section{Gross-Neveu-Yukawa model in $d=4-2\epsilon$ dimensions}\label{GNY}
The Lagrangian of the GNY model in $d=4-2\epsilon$ Euclidean space takes the form
\begin{align}
\mathcal{L}=
\frac12(\partial\sigma)^2+\bar q\slashed{\partial} q+g_1 \sigma \bar q q+ g_2\sigma^4,
\end{align}
where $q$ is the $N$-component fermion field and $\sigma$ is a scalar field.
 The model is multiplicatively renormalized~\cite{ZinnJustin:1991yn}
\begin{align}\label{eq:ren_action}
 \mathcal{L}_R = \frac{Z_1}2(\partial\sigma)^2 +Z_2\bar q\slashed{\partial}q
+M^\epsilon Z_3 g_1 \sigma \bar q  q
+M^{2\epsilon}Z_4 g_2\sigma^4.
\end{align}
Here $M$ is the renormalization scale and $Z_i$ are the renormalization constants. In MS-like schemes the renormalization
factors $Z_i$ are given by a series in
$1/\epsilon$,
$Z_i=1+\sum_k z_{i}^k/\epsilon^k$.

The anomalous dimensions of the fields and the $\beta$ functions are defined as usual
\begin{align}
  \gamma_q =\mathcal{ D}_M \ln Z_q, &&
  \gamma_\sigma = \mathcal{D}_M \ln Z_\sigma, &&\beta_i=\mathcal{ D}_M g_i,
\end{align}
where  $Z_\sigma= Z_1^{1/2}$ and $Z_q = Z_2^{1/2}$ and
\begin{align}\label{DM}
\mathcal{D}_M \equiv M\frac{d}{dM}=M\partial_M+\beta_1\partial_{g_1}+\beta_2\partial_{g_2}\,.
\end{align}
%
The GNY model  possesses non-trivial  fixed points 
\begin{align}
  \beta_1(g_1^\ast,g_2^\ast) = 0 , &&
  \beta_2(g_1^\ast,g_2^\ast) = 0.
\end{align}
One of this points is infrared stable, i.e. the eigenvalues of the matrix
\begin{align}
\omega_{ik}=\partial_i \beta_k\big|_{g=g^\ast}
\end{align}
are both positive. These eigenvalues, $\omega_\pm$, 
are called correction indices. The indices $\omega_\pm$ can be identified with the anomalous dimensions of certain composite
operators that will be discussed in the rest of this section.

The partition function $Z(J)$ defined by the path integral
\begin{align}\label{ZJ}
Z(J)=\int D\bar q Dq D\sigma \exp\left\{-S_R+ \int d^dx\,\, J(x)\,\Phi(x)\right\},
\end{align}
where $J=\{j,\eta,\bar \eta\}$ and $J\Phi=j\sigma+\bar\eta q+\bar q\eta$,  is  finite in perturbation theory. Moreover,
derivatives of \eqref{ZJ} with respect to the renormalized couplings $g_i$  are finite, too. Thus one concludes that
\begin{align}
\partial_{g_k} \mathcal{L}_R= \sum_a c_a [\mathcal{O}_a]+ \partial \mathcal{ B}\,,
\end{align}
where the sum goes over some set of renormalized operators of the canonical dimension four with finite coefficients~$c_a$. By
$\partial \mathcal{ B}$ we denote terms (not necessarily finite) which vanish after integration. Since
\begin{align}
\partial_{g_1} \mathcal{L}_R \equiv M^{\epsilon} \sigma \bar q q + \text{singular $1/\epsilon$ terms}
\end{align}
and similar for another derivative one concludes that
\begin{align} \label{eq:der_action_ren}
\partial_{g_1}\mathcal{L}_R & = M^{\epsilon} [\sigma \bar q q]\equiv M^\epsilon[\mathcal{O}_1] ,
\notag\\
\partial_{g_2} \mathcal{L}_R &= M^{2\epsilon} [\sigma^4]\equiv M^{2\epsilon} [\mathcal{O}_2],
\end{align}
where we omitted terms with total derivatives.

Applying the operator~\eqref{DM}  to both sides of Eqs. \eqref{eq:der_action_ren} and taking into account that
$\mathcal{D}_M S_R=\mathcal{D}_M S_0=0$ one derives the RG equation for the operators $[\mathcal{O}_k]$
\begin{align} \label{eq:RGE}
\Big(\delta_{ik}\mathcal{D}_M  +\gamma_{ik}\Big)[\mathcal{O}_k] =0\,,
\end{align}
where the anomalous dimension matrix $\gamma$ has the following form
\begin{align}\label{eq:gammacrit}
  \gamma =
  \begin{pmatrix}
    \epsilon + \partial_{g_1} \beta_{1} & M^{\epsilon} \partial_{g_1} \beta_{2} \\
     M^{-\epsilon} \partial_{g_2} \beta_{1} & 2\epsilon + \partial_{g_2} \beta_2
  \end{pmatrix}.
\end{align}
Using the known lowest order expression for  the beta-functions (see e.g. ref.~\cite{Zerf:2017zqi}) one finds that at the
Wilson-Fisher critical point  the anomalous dimensions take the form
\begin{align}
\gamma =
  \begin{pmatrix}
    3\epsilon & 0 \\
    0 & 4\epsilon
  \end{pmatrix} +O(1/N).
\end{align}
resulting in the following  scaling dimensions for  the operators $\mathcal{O}_\pm$,
\begin{align}
\Delta & =\Delta_\text{can}+\gamma_\ast
\notag\\
& = d+\begin{pmatrix}
    \partial_{g_1} \beta_{1} & M^{\epsilon} \partial_{g_1} \beta_{2} \\
     M^{-\epsilon} \partial_{g_2} \beta_{1} &  \partial_{g_2} \beta_2
  \end{pmatrix}=4+O(1/N).
\end{align}
Diagonalizing the matrix $\Delta$ one constructs two operators whose scaling dimensions are given  by the eigenvalues
of~$\Delta$. Scaling dimensions of  operators are physical observables and do not depend on a regularization, expansion,
subtraction scheme, etc. We use this property to calculate the anomalous dimension of the operators $[\mathcal{O}_\pm]$  in
$1/N$ expansion.

\section{Large $N$ expansion for the GN model}\label{largeN}
The GNY model is critically equivalent to the  $d$-dimensional $N$-component GN model, i.e.  critical indices in both models
coincide. In the GN model the indices can be calculated using the $1/N$ expansion. Below we briefly review the method that we
use for the calculation. A more extensive review of the $1/N$ techniques including the self-consistency equations and conformal
bootstrap can be found in the book~\cite{Vasilev:2004yr}.

Let us write the action of the GN model in the form suitable for generating
the $1/N$ expansion
\begin{align}\label{N-GN}
S_{GN} &=\int d^d x\left[\bar q\slashed{\partial} q+\sigma \bar q q -\frac{N}{2g} \sigma^2\right]\,.
\end{align}
Here $q=\{q^i, i=1,\ldots,N\}$ is the $N$-component fermion field and $\sigma$ is an auxiliary scalar field which can be
excluded by the equation of motion (EOM). At a certain (critical) value of the coupling, $g=g_*$ the system undergoes a second order phase
transition. At  this point correlators of the basic and auxiliary fields
$q,\bar q, \sigma$ exhibit  power law behaviour and, as it can be shown, the model enjoys  scale and conformal
invariance~\cite{Derkachov:1993uw}.

It can be shown that in the {\it infrared region} (IR) the dominant contribution to the propagator of the $\sigma$-field comes
from the fermion loop~\cite{ZinnJustin:1991yn}
\begin{align}\label{sigma-prop}
D_\sigma(x)=-\frac1n {B(\mu)}/{x^2}\,.
\end{align}
Here $n=N\,\text{tr} \II$, where the trace is taken in the space of $d$-dimensional spinors
 and the normalization factor is
\begin{align}\label{bmu}
B(\mu)=\frac{4\Gamma(2\mu-1)}{\Gamma^2(\mu)\Gamma(\mu-1)\Gamma(1-\mu)}\,.
\end{align}
For practical calculations   it is convenient to use  a simplified (massless) version of the GN model which is critically
equivalent to~\eqref{N-GN}. The action of the model is given by the following expression~\cite{Vasiliev:1975mq,Vasilev:2004yr}
\begin{align}\label{L-GN}
S'_\Delta &=\int d^d x\left[\bar q\slashed{\partial} q -\frac12 \sigma L_\Delta\sigma +\sigma \bar q q  + \frac12 \sigma L\sigma\right]\,.
\end{align}
The kernel $L$
has the form
\begin{align}
L(x) & = \mathrm{tr} D_q(x) D_{q}(-x)
=-n
{A^2(\mu)}/{x^{2(2\mu-1)}}\,,
\end{align}
where $D_q$ is the fermion propagator
\begin{align}
D_q(x) =-
\frac{A(\mu)\slashed{x}}{x^{2\mu}} \,, &&A(\mu)=\frac{\Gamma(\mu)}{2\pi^\mu}\,.
\end{align}
The regularized kernel $L_\Delta$ is
\begin{align}\label{Ldef}
L_\Delta(x)= L(x) (M^2 x^2)^{-\Delta} C^{-1}(\Delta)\sim x^{-2(2\mu-1+\Delta)}\,.
\end{align}

The first two terms in~\eqref{L-GN} can be viewed as the free part of the action, $S_0$, and the remaining ones -- as an
interaction.
The last term in~\eqref{L-GN} cancels diagrams with insertions of simple fermion loops in the
$\sigma$-lines. As a consequence, in the leading order the propagator of the $\sigma$ field
is given by the inverse kernel $L_\Delta$. It this work we fix the constant $C(\Delta)$, which is arbitrary save the condition
$C(0)=1$, by the requirement for the propagator $D_\sigma$ to have the following form
\begin{align}
D_\sigma(x)=-  \frac1{n} B(\mu)(M^{2}x^2)^\Delta/{x^2}\,.
\end{align}
The parameter $\Delta$ should be considered as a regularization parameter. The divergences in the correlators appear as
poles in $\Delta$ and are removed by adding the counterterms to the action~\eqref{L-GN}. The renormalized action takes the form
\begin{align}\label{L-GN-ren}
S'_{R} &=\int d^d x\left[Z_1\bar q\slashed{\partial} q -\frac12 \sigma L_\Delta\sigma + Z_2 \sigma \bar q q
+ \frac12 \sigma L\sigma\right]\,.
\end{align}
Though the model is renormalizable, the renormalization is not multiplicative, i.e.  $S'_R(q,\sigma) \neq S'(q_0,\sigma_0)$.
Due to the non-multiplicative character  of the renormalization  the anomalous dimensions of the fields and operators cannot
be, in general, determined via the corresponding renormalization factors. However, as it was shown in~\cite{Derkachov:1997ch}
this problem appears  starting from
$1/n^3$ order only. Up to the order $1/n^2$ the anomalous dimension of the operators can be obtained via the renormalization
factors as follows. First, one has to modify the $\sigma$-propagator by multiplying it by some parameter $u$,
$D_\sigma \mapsto u D_\sigma$ and determine the renormalization factors calculating  the corresponding diagrams with the modified propagator.
Let $\mathcal{O}_i$ be a system of operators mixing under renormalization,
\begin{align}
[\mathcal{O}_i(\Phi)] =(\mathbf{Z}(u))_{ik} \mathcal{O}^B_k(\Phi_0)\,,
&&
 \mathbf{Z}(u)=1+\sum_{a=1}^\infty \frac{Z_a(u)}{\Delta^a}\,,
\end{align}
where $\Phi=\{q,\bar q, \sigma\}$, and $\mathcal{O}^B_i$ are bare operators. Up to the order $1/n^2$ the anomalous
dimensions for the operators can be obtained as follows~\cite{Derkachov:1997ch}
\begin{align}\label{gamma-O}
\gamma_{ik}= 2\partial_u (\mathbf{Z}_1(u))_{ik}\Big|_{u=1} +O(1/n^3)\,.
\end{align}
We use this formula for the calculation of the anomalous dimensions of the operators in question.

\section{Correction exponents at $1/n^2$}\label{diagrams}
\begin{figure}[t]
  \centering
  \includegraphics[width=0.5\columnwidth]{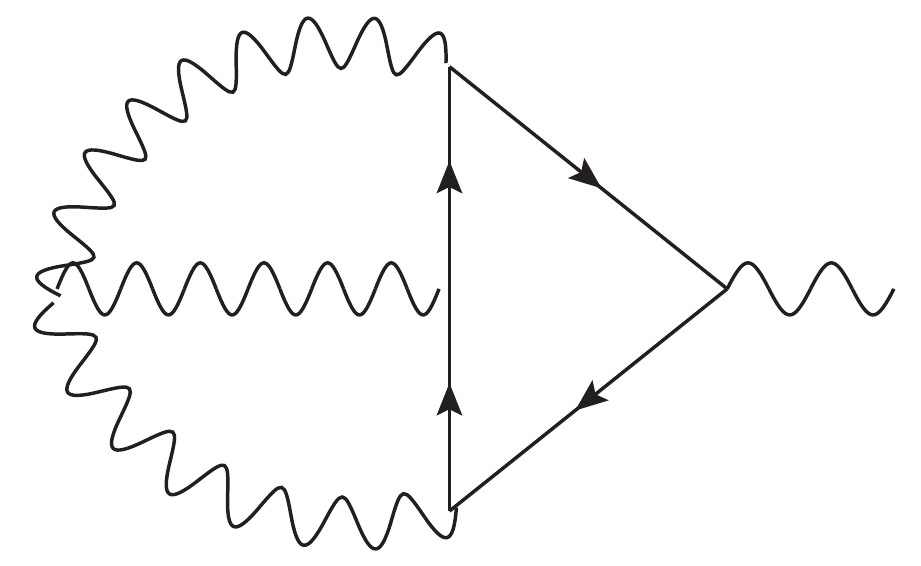}
  \caption{The $1/n^2$ diagram $\sigma^3\mapsto \sigma\partial^2\sigma$ which could contribute to the element $\gamma_{21}$ of the mixing
  matrix. }
  \label{fig:Mixing}
\end{figure}
In the perturbative expansion the correction exponents to the scaling are related to the scaling dimensions of the operators
$\sigma^4$ and
$\sigma \bar q q$. The scaling dimensions of these operators are $\Delta_a=4+O(1/n)$. Let us first identify the corresponding
operators in the $1/n$ expansion. There exist three scalar operators of dimension four,
$$
\mathcal{O}=\{\sigma^4,\sigma\partial^2\sigma,  \partial^2 \sigma^2\}.
$$
\begin{figure*}[t]
\begin{minipage}{\linewidth}
\centering
  \includegraphics[width=0.52\linewidth]{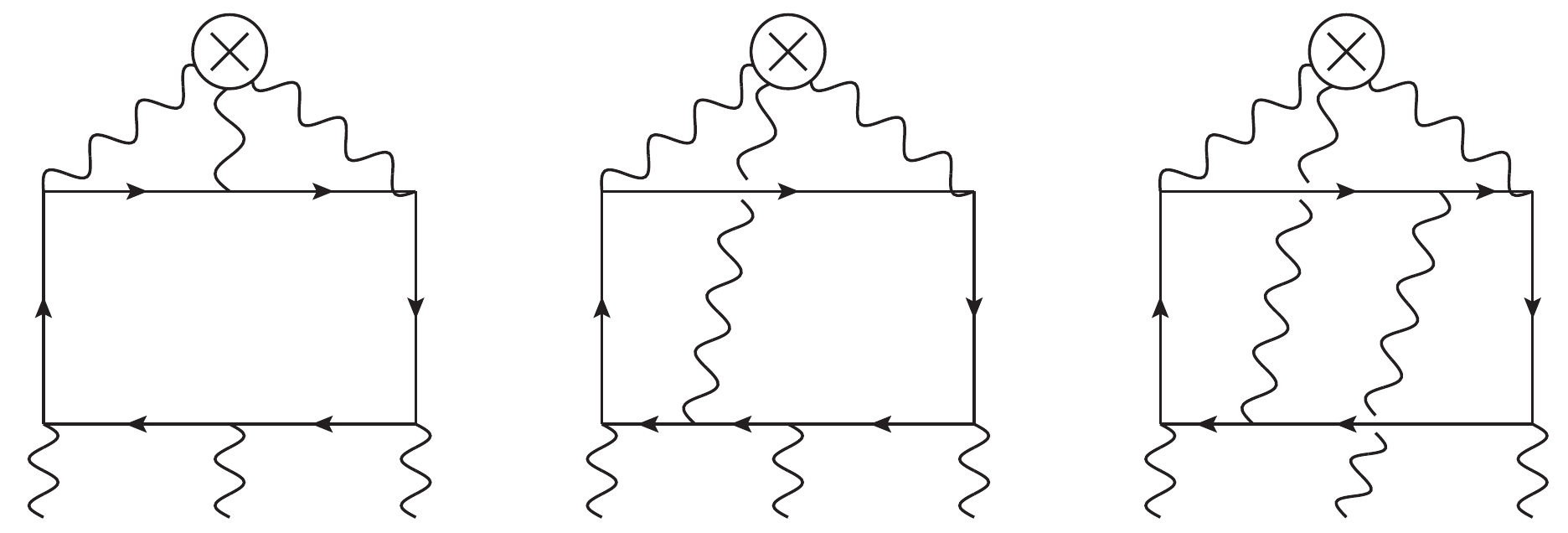}\\[3mm]
  \includegraphics[width=0.99\linewidth]{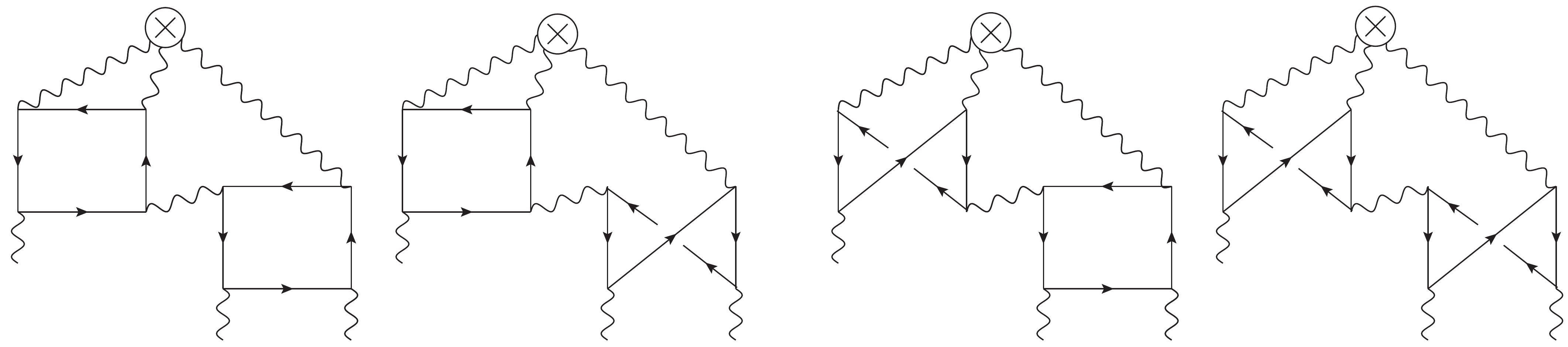}
  \caption{Feynman diagrams $D_{1}- D_3$ (first line) and $D_4- D_7$ (second line)
contributing to the anomalous dimensions $\Delta \gamma_{\sigma^3}$ }
  \label{fig:Dgamma3}
  \end{minipage}
\end{figure*}
 The last one is a total derivative and can be
neglected. Note here that the operator $\sigma\bar q q$ in the $1/n$ expansion has the dimension $4-2\epsilon+O(1/n)$.
Therefore it does not mix with the above operators and has no relation to the correction exponents~\footnote{The universality
hypothesis states that theories in the same universality class possess the same spectrum of  scaling dimensions, but the
corresponding operators can have  different implementation.}. Thus we need to find the anomalous dimension matrix for the
operators
$\mathcal{O}_+=\sigma^4$ and $\mathcal{O}_-={\sigma\partial^2\sigma}$.

It is easy to see that
 the diagonal entries $\gamma_{11}$, $\gamma_{22}$ and $\gamma_{12}$ of the anomalous dimension matrix are of order $1/n$. At the same time the anomalous
 dimension $\gamma_{21}$ is of order $1/n^3$. It happens because the only diagram which contributes  to this matrix element at order
 $1/n^2$, $\sigma^3\mapsto \sigma\partial^2\sigma$, see Fig.~\ref{fig:Mixing},
does not diverge. It can be checked by an explicit calculation.
 Also, the finiteness of this diagram is guaranteed by the conformal symmetry. The arguments are exactly the same as in the case of
 $\sigma^2\mapsto \partial^2\sigma$ transition in the nonlinear $\sigma$ model. An interested reader can find detailed
 discussions in  Refs.~\cite{Derkachov:1997gc,Derkachov:1998js}. Thus at the order $1/n^2$ one can neglect mixing and
 calculate only the diagonal matrix elements.

We will use either upper or lower index for  the coefficients  of the $1/n$ expansion
\begin{align}
\gamma=\sum_{k=0}^\infty \gamma_k/n^k=
\sum_{k=0}^\infty \gamma^{(k)}/n^k
\end{align}
  trying  to keep  notations close to   generally
accepted. For the reader's convenience we collect $1/n^2$ expressions
for the basic indices ($\eta=2\gamma_q$,
$\gamma_\sigma=-\eta-\kappa$ and $\lambda=(d-\Delta[\sigma^2])/2$) in \ref{app:1/n^2indices}.

\subsection{Operator $\sigma^\ell$}

Here we present our result for the anomalous dimension of the operator $\sigma^\ell$. The mixing  affects the anomalous
dimension of this operator only starting from $1/n^3$. Analyzing  the contributing diagrams  it is easy to notice that
the anomalous dimension for the operator
$\sigma^\ell$ can be written in the form
\begin{align}
\gamma_{\sigma^\ell}=-\ell(\ell-2)\gamma_\sigma+ 
C^\ell_2\gamma_{\sigma^2} + 
C^\ell_3\Delta\gamma_{\sigma^3}+O(1/n^3)\,,
\end{align}
where $C^\ell_k$ are the binomial coefficients. The anomalous dimension $\gamma_\sigma$ and $\gamma_{\sigma^2}$ are known with
$1/n^2$ accuracy~\cite{Gracey:1992cp,Gracey:1990wi,Derkachov:1993uw,Vasiliev:1993pi}, see \ref{app:1/n^2indices}. The last
contribution starts from $1/n^2$. The diagrams contributing to
$\Delta\gamma_{\sigma^3}$ are shown in Fig. \ref{fig:Dgamma3}.
Their calculation  is rather straightforward so we will not dwell on it. We obtained  the following expression for
$\Delta\gamma_{\sigma^3}$
\begin{align}
  \Delta \gamma_{\sigma^3}^{(2)} =
\eta_1^{2}\frac{6\mu^2}{\mu-1}\Big[&-\frac{(3\mu-2)(2\mu-1)}{\mu-1}\notag
\\
&+(3\mu-1)\big[\psi'(\mu)-\psi'(1)\big]
\Big]\,.
\end{align}
Here $\psi(x)$ is the Euler $\psi$ function and an explicit expression for the index $\eta_1=-B(\mu)/2\mu$ can be found in
\ref{app:1/n^2indices}.  Thus for the correction index $\omega_+$ we find
\begin{align}\label{eq:omega+}
\omega_+&=2\Big(\epsilon-4\gamma_\sigma + 3\gamma_{\sigma^3}+2\Delta \gamma_{\sigma^3}+O(1/n^3)\Big)
\notag\\
&=2\epsilon +\frac{\eta_1}{n}\frac{4(2\mu-1)(3\mu-1)}{\mu-1}+\ldots
\end{align}
The explicit expression for the $1/n^2$ correction, $\omega_+^{(2)}$, is rather lengthy and will not be given here. Instead we
present an expansion of the index $\omega_+$
around $\mu =2-\epsilon$ up to  $\epsilon^4$ terms
\begin{align}
\omega_+^{(0)}&=2\epsilon\,,
\notag\\
\omega_+^{(1)}&= 120\epsilon-212\epsilon^2-26\epsilon^3 +\big(-29+240 \zeta_3\big)\epsilon^4\,,
\notag\\
\omega_+^{(2)}&=-5040\epsilon+7380\epsilon^2 + 2\big(8767+6624\zeta_3\big)\epsilon^3
\notag\\
&\quad +
32\left(-392 + 621 \zeta_4 - 2226\zeta_3 - 420\zeta_5\right)\epsilon^4\,,
\end{align}
in complete agreement with the results of the four--loop calculation~\cite{Zerf:2017zqi}.

\subsection{Operator $\sigma\partial^2\sigma$}
\begin{figure}[H]
  \centering
  \includegraphics[width=0.7\linewidth]{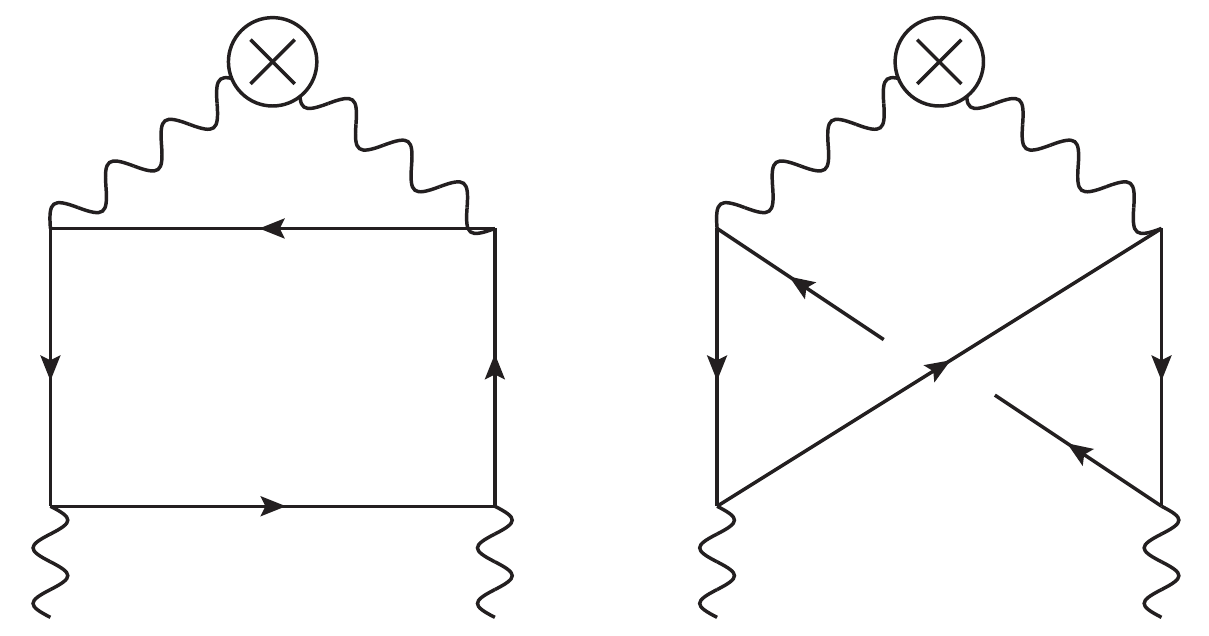}
  \caption{ The LO diagrams contributing
to the anomalous dimension~$\gamma_{-}$. The circled cross denotes
an operator insertion $\mathcal O_-$. }
  \label{fig:g-LO}
\end{figure}

In this section we discuss diagrams contributing to the anomalous dimension of  the operator $\mathcal O_-= \sigma
\partial^2\sigma$ which, as usual, can be split in two parts
\begin{align}
\gamma_-=\widehat \gamma +2\gamma_\sigma\,.
\end{align}
In the leading order in $1/n$ only the two diagrams shown in Fig.~\ref{fig:g-LO} contribute to
$\widehat \gamma$.
\noindent The first diagram has the symmetry coefficient $C_{\text{sym}}=2$. The calculation of both diagrams is rather
straightforward and leads to the following expression
\begin{gather}
\widehat \gamma_1 =   2 \eta_1 
{(4 \alpha^2- 1)}/{\alpha},
\end{gather}
where we introduced the notation $\alpha \equiv \mu-1$. Taking into account that $\gamma_\sigma^{(1)}=-\eta_1(1+\mu/\alpha)$
one obtains for
$\gamma_-$
\begin{align}
\gamma_-^{(1)}=\eta_1 \frac{4(2\mu-1)(\mu-2)}{\mu-1}
\end{align}
that agrees  with the result of~\cite{Gracey:2017fzu} derived with the help of the  self-consistency equations approach.

\begin{figure}[H]
  \centering
  \includegraphics[width=\columnwidth]{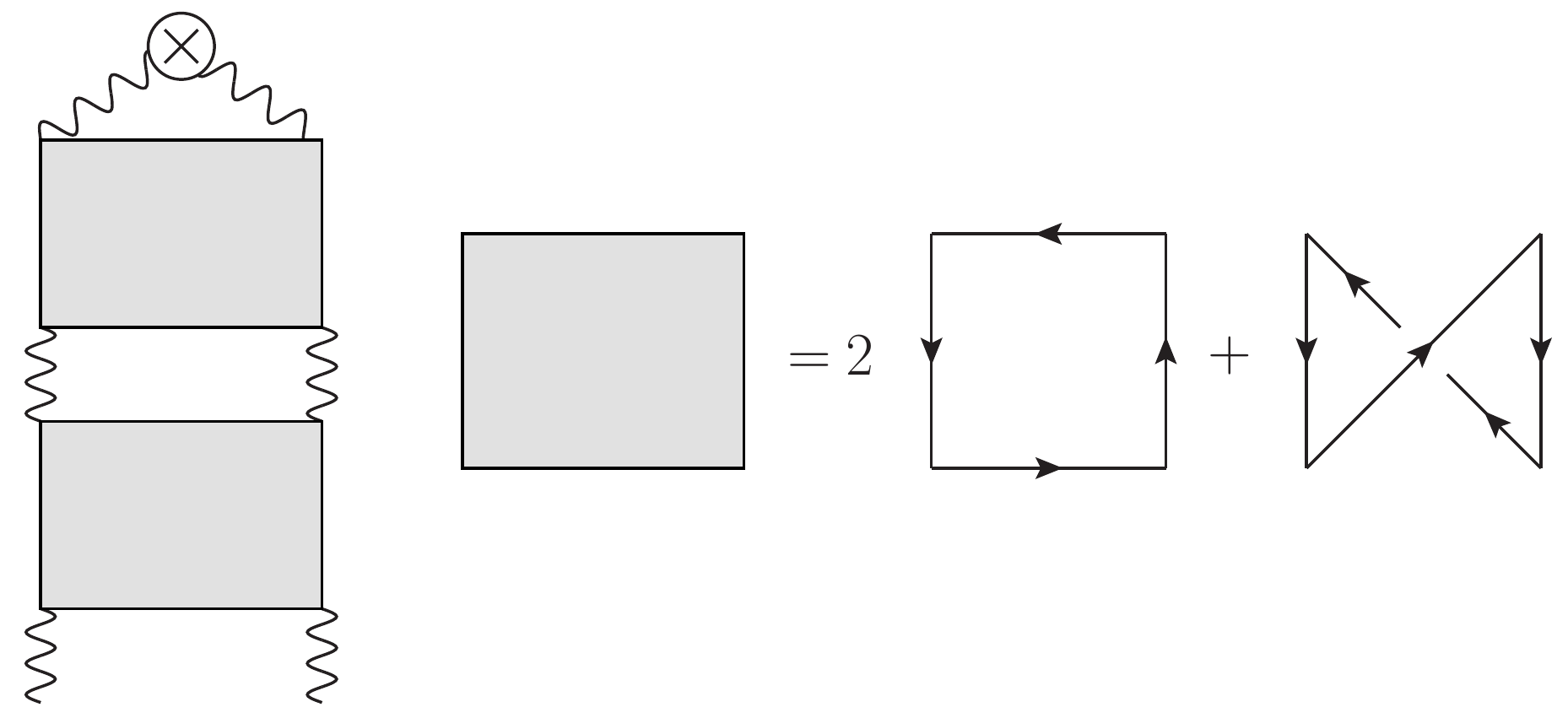}
  \caption{ Operator vertex correction diagrams.
}
  \label{fig:OC}
\end{figure}

At the  order $1/n^2$  there are in total $23$ different diagrams to be calculated. Namely,
\begin{itemize}
\item
    $4$ operator vertex correction diagrams, Fig.~\ref{fig:OC}
\item
    $12$ self-energy insertion and vertex correction diagrams
\item
    $3$~single box diagrams, Fig.~\ref{fig:SingleBox}
\item
    $4$ double box diagrams, Fig. \ref{fig:DoubleBoxes}
\end{itemize}
Before presenting our final answer for $\widehat \gamma$ we briefly discuss the calculation of the diagrams.
\begin{itemize}
\item In order to calculate the operator correction diagrams it is sufficient to calculate
the $1/n$ order diagrams up to
    finite,
$O(\Delta^0)$, terms.  For the sum of the two diagrams we obtained~\footnote{We do not display  factors like
$t^\Delta=e^{-\Delta \psi(1)+\ldots}$ which drop out from the final answer.}
 $\mathbb D (\Delta)= u^2 
 D(\Delta)/\Delta$, where
\begin{align}
 D(\Delta) &=-\frac{\widehat \gamma_1}{4}
\Biggl\{
1-\Delta\biggl[2B_3 -\frac{3\mu(\mu-1)(\mu-2)}{(2\mu-1)(2\mu-3)} C_1
\notag \\
&\quad
+ \frac{\mu^5-9\mu^4+21\mu^3-9\mu^2-6\mu+4}{(2\mu-1)\mu(\mu-1)(\mu-2)}
\biggr]
\Biggr\}\,,
\end{align}
where  we accepted the notations of~\cite{Vasilev:2004yr}
\begin{align}
B(x) &=\psi(x)+\psi(x')\,, &&
C_x=\psi'(x)-\psi'(x'),
\end{align}
$ B_k=B(k-\mu)-B(1)$ and $x'\equiv \mu-x$.

The answer for the operator correction diagrams ($D_\text{OC}) $ reads
\begin{align}
KR'(D_\text{OC}) &= u^4\left( 
\frac{D(\Delta)}{\Delta}\frac{ D(2\Delta)}{2\Delta}-
\frac{D(0)D(\Delta)}{\Delta^2}\right)
\notag\\
&=\frac12 u^4 D^2(0)\left(-\frac1{\Delta^2}+\frac1{\Delta} \frac{D'(0)}{D(0)}\right)\,.
\end{align}
\item Instead of calculating separately  each diagram of the second group, that is not quite simple, we use the result
    of~\cite{Derkachov:1997ch} where it was shown that the contribution of all such diagrams to the anomalous dimension can
    be extracted from the $1/n$ order diagrams with dressed propagators and vertices. We   discuss  details of the
    calculation in \ref{app:SEV} .

\item The single box diagrams shown in Fig.~\ref{fig:SingleBox} as well as double box diagrams, Fig. \ref{fig:DoubleBoxes},
    have only a superficial divergence.
\begin{figure}[H]
  \centering
  \includegraphics[width=0.98\columnwidth]{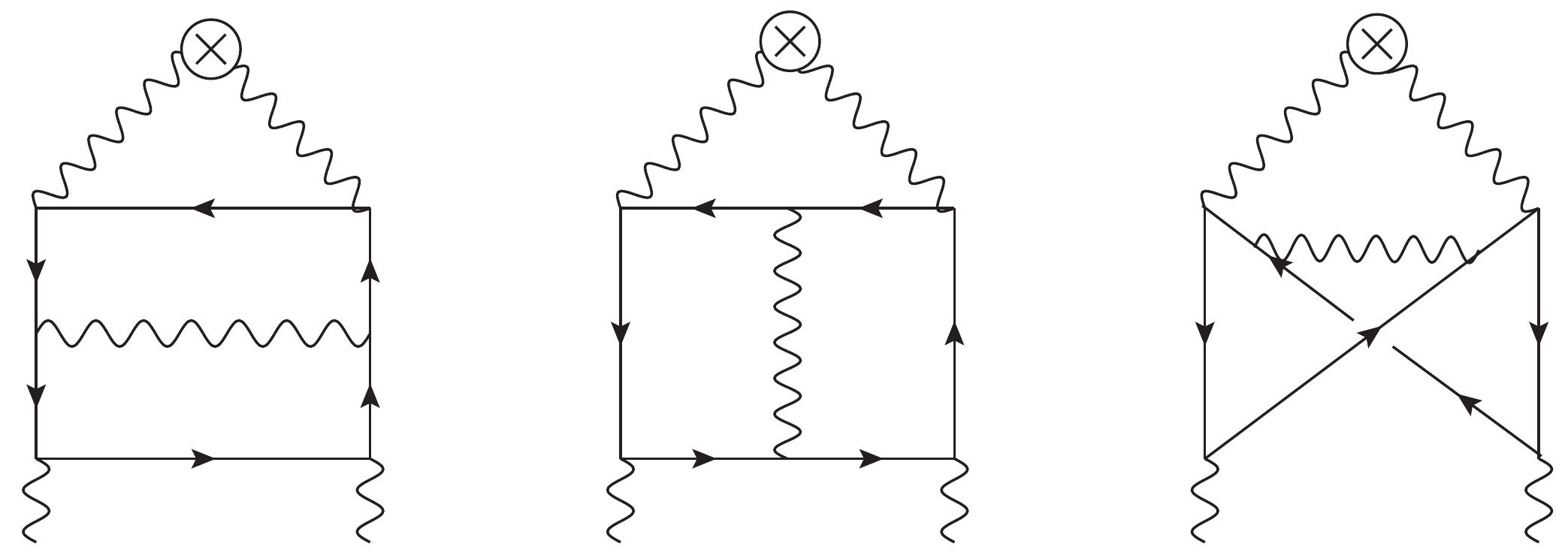}
  \caption{ Single box diagrams $SB_1$ - $SB_3$.
}
  \label{fig:SingleBox}
\end{figure}
\vskip -5mm After integration over the operator insertion coordinate one gets a propagator  type diagram,
    \begin{align}
    D_k(x)= A_k(\Delta)/x^{2(\mu+1-3\Delta)}\,.
    \end{align}
The $\Delta$ pole arises after the Fourier transform. For the anomalous dimension it is sufficient to know $A_k(\Delta)$ at
$\Delta=0$. Since in coordinate space the diagram is finite at $\Delta\mapsto 0$, one can put $\Delta=0$ from the beginning.
It greatly simplifies  calculations since one can use the star~-~triangle relation for the basic  $\sigma \bar q q$ vertex,
shown in Fig.~\ref{fig:star-triangle}.
\begin{figure}[H]
  \centering
  \includegraphics[width=0.92\columnwidth]{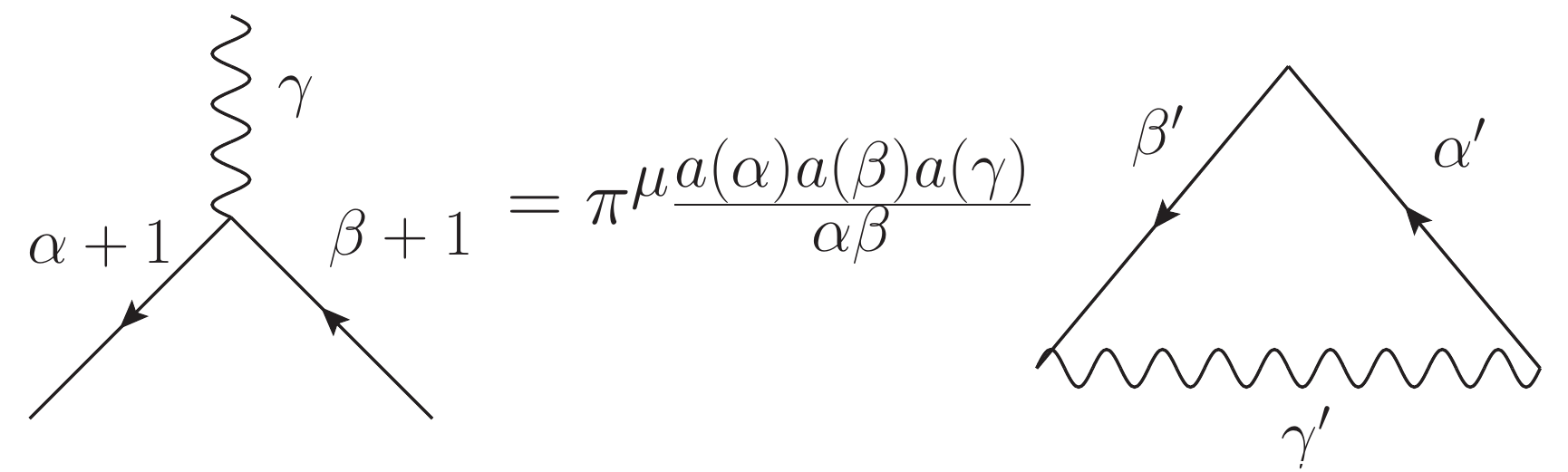}
  \caption{ The star-triangle relation. The indices $\alpha,\beta,\gamma$ satisfy the uniqueness condition
  $\alpha+\beta+\gamma=2\mu-1$, $a(x)=\Gamma(x')/\Gamma(x)$ and $x'=\mu-x$.
}
  \label{fig:star-triangle}
\end{figure}
 Next, multiplication by $x^2$ makes the diagrams logarithmically divergent, i.e. the residue of the $\Delta$ pole does not
depend on the external momenta. Thus one can use the freedom to change the momentum flow through a diagram to facilitate
calculations~\cite{Vladimirov:1979zm}. For example, to calculate the second  diagram in Fig.~\ref{fig:SingleBox} it is
convenient to choose the momentum flow (after a multiplication by $x^2$) through the vertices connected to the crossed
circle (operator insertion) and use the star-triangle relation for the upper $\sigma \bar q q$ vertex. After this
transformation the diagram is calculated by consecutive application of chain integration rules.

\item The calculation of the first three double box diagrams which follows the same pattern is more involved and one has to
    use many tricks like the integration by part~\cite{Chetyrkin:1981qh}, the tetrahedron
    transformation~\cite{Gorishnii:1984te}, etc. A review of the relevant multiloop calculation techniques can be found
    in~\cite{Vasilev:2004yr}. \vskip-2mm
\begin{figure}[H]
  \centering
  \includegraphics[width=0.98\columnwidth]{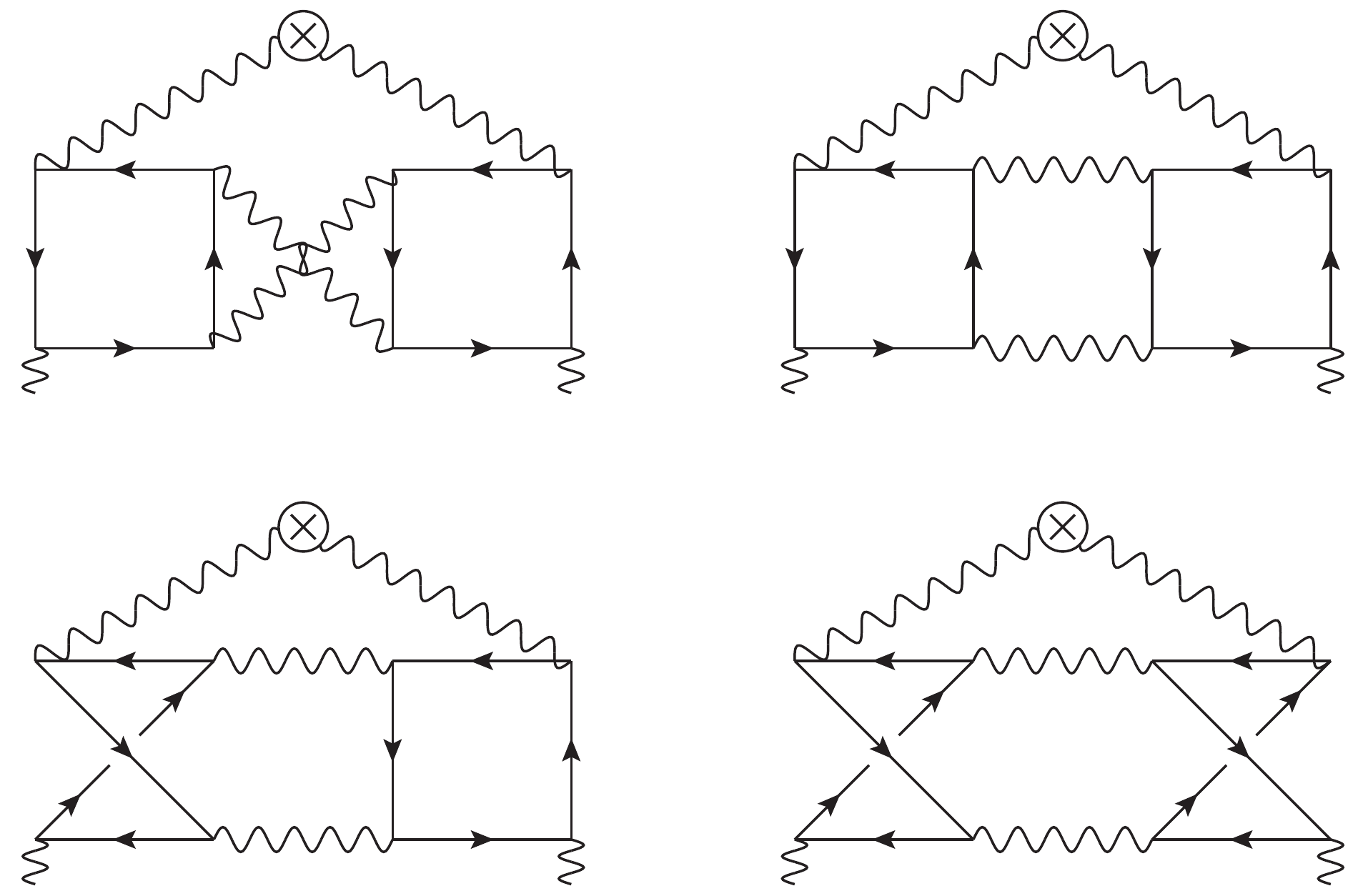}
  \caption{ Double box diagrams $DB_1$ (top left) - $DB_4$ (bottom right).
}
  \label{fig:DoubleBoxes}
\end{figure}
\vskip -3mm In principle the last double box diagram could be calculated in the same way. However it is more convenient
 to apply an inversion transformation first, simplify the numerators and use the star-triangle relation for
the bosonic triangle. As the result one gets
a sum of three- and two-loop diagrams whose calculation is more or less straightforward.

\end{itemize}

Finally, collecting contributions from  all diagrams (see ~\ref{app:part_results}) we obtain for
$\widehat\gamma_2$
 {\allowdisplaybreaks
\begin{align}
 \widehat \gamma_2 = &
\eta_1 \Biggl\{
-\frac{8 (2 \alpha -1) \left(2 \alpha ^3+5 \alpha -2\right)}{(\alpha -2) (\alpha -1) \alpha} \notag
\\
  &+\eta_1\biggl(
  -
  {3 \left(4 \alpha ^3+2 \alpha ^2+1 -1/\alpha\right)}
  C_1
 \notag \\
  &+\left[32 \alpha ^2+\frac{2}{\alpha ^2}+40 \alpha +\frac{36}{\alpha -2}+\frac{20}{\alpha
   -1}-\frac{6}{\alpha }+40\right]B_3 \notag \\
  &+8 \alpha ^3+\frac{4}{\alpha ^3}-8 \alpha ^2-\frac{12}{\alpha ^2}-14 \alpha
   -\frac{18}{\alpha -2}\notag \\
&+\frac{72}{\alpha -1}-\frac{18}{\alpha +1}+\frac{20}{(\alpha-1)^2}+\frac{10}{\alpha }+84
  \biggr)
\Biggr\}.
\end{align}}
 Thus for the second correction exponent we get
 \begin{align}
 \omega_-  =2\epsilon+\gamma_- +O(1/n^3)=2\epsilon +2\gamma_\sigma+\widehat\gamma+ O(1/n^3)\,.
 \end{align}
Substituting the expressions for $\gamma_\sigma$ and $\widehat\gamma$ and expanding around $\mu=2-\epsilon$ up to $\epsilon^4$
terms one obtains
\begin{align}
\omega_-^{(0)} =&2\epsilon\,,
\notag\\
\omega_-^{(1)} =&-24 \epsilon ^2+28 \epsilon ^3+22 \epsilon ^4\,,
\notag\\
\omega_-^{(2)} =& 174 \epsilon ^2+\left(432 \zeta_3-293\right)\epsilon ^3   \notag \\
&-2\left(1500 \zeta_3-324 \zeta_4+419\right)  \epsilon ^4\,,
\end{align}
that agrees with the results of \cite{Zerf:2017zqi}.

\section{Summary}\label{summary}
We have calculated the correction exponents $\omega_\pm$ in the GN model with $1/n^2$ accuracy. These exponents are related to
the slopes of the beta functions in the critical GNY model. Our results are in complete agreement with the expressions for the
perturbative  four loop beta functions in the GNY model obtained recently in~\cite{Zerf:2017zqi}. For the calculation we used
the method developed in~\cite{Vasiliev:1975mq,Vasiliev:1993ux,Derkachov:1997ch}. This method allows one to use the standard RG
technique and, at the same time, to resum effectively a certain subset of diagrams reducing the total number of diagrams to be
calculated. Values for each individual diagram are presented in~\ref{app:part_results} and will be useful for
calculations of the corresponding exponents  in  the chiral versions of the GN model.

\begin{acknowledgements}
This work was supported by Deutsche Forschungsgemeinschaft (DFG) with the grants
$\text{MO~1801/1-2}$ (A.M.) and SFB/TRR 55 (M.S.)
\end{acknowledgements}

\appendix
\section{$1/n^2$ indices}\label{app:1/n^2indices}
In this appendix we collected the results for  the basic indices $\eta$, $\kappa=2\chi$ and~$\lambda$ related to the anomalous
dimensions of the  fields $q$, $\sigma$ and the composite operator
$\sigma^2$
\begin{align}\label{indices}
  \gamma_q = \eta/2 \,,
&&
  \gamma_{\sigma}= 
  -\eta-2\chi\,,
&&
  \gamma_{\sigma^2} = -2\lambda\,,
\end{align}
that can be found either in the original
papers~\cite{Gracey:1990wi,Gracey:1992cp,Derkachov:1993uw,Vasiliev:1993pi,Gracey:1993kb} or  in the
book~\cite{Vasilev:2004yr}.
The leading  order coefficients 
read
\begin{align}
  \eta_1 &=  -\frac{2\Gamma(2\mu-1)}{\Gamma(\mu)\Gamma(\mu+1)\Gamma(\mu-1)
\Gamma(1-\mu)} \,, \notag \\[2mm]
  \kappa_1 &=\eta _1 
  { \mu }/(\mu-1) \,,
  \notag\\[2mm]
  \lambda_1 & = -\eta_1(2\mu-1)\,
\end{align}
and the $1/n^2$  coefficients take the form
\begin{align}
  \eta_2 =& \eta _1^2 \left(\frac{1}{2 \mu }-\frac{\mu}{2 (\mu -1)^2}
+\frac{(2 \mu-1)}{\mu -1}B_1\right), \notag \\
\kappa_2 =&
\eta _1^2\frac{\mu}{\alpha}  \bigg(\frac{2 \mu -1}{\alpha}B_1
-3 \mu  C_1-2 \mu -\frac{6}{\alpha}-\frac{4}{\alpha^2}-1\bigg)\,,
\notag \\
  \lambda_2 =& \eta _1\frac{\mu}{2\alpha} \bigg\{\frac{8}{(2-\mu)^2}
+\eta_1\bigg[
\frac{4 (3-2 \mu ) \mu  \left(C_{2-\mu }-B_2^2\right)}{2-\mu }
\notag \\
&-\left(8 \mu^2+\frac{10}{2-\mu }+\frac{2}{\mu }-\frac{4}{(2-\mu)^2}
+\frac{8}{\alpha}-2\right) B_2 \notag \\
&+ \mu\left(6 \alpha+\frac{22}{2-\mu }-29\right)C_1  -\frac{1}{\mu ^2}+8 \alpha^2
\notag \\
&+\frac{42}{2-\mu }+\frac{4}{\mu}-\frac{4}{(2-\mu )^2}+\frac{14}{\alpha}-\frac{5}{\alpha^2}-10
\bigg]
\bigg\} \,.
\end{align}
We remind here that $\alpha=\mu-1$, $\eta= \sum_i \eta_i/n^i$, etc.
\section{SE and Vertex correction diagrams}\label{app:SEV}
\begin{figure}[H]
  \centering
  \includegraphics[width=0.98\columnwidth]{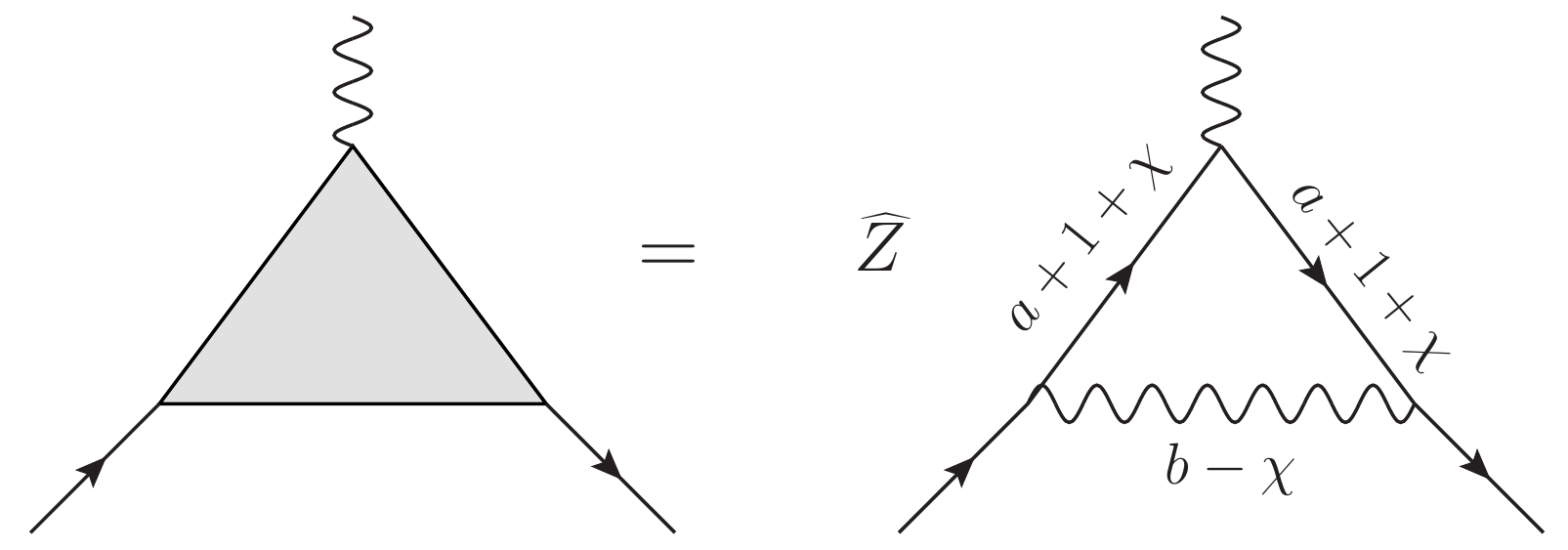}
  \caption{ Dressed $\sigma \bar q q$ vertex.}
  \label{fig:Vertex}
\end{figure}

It was shown in~\cite{Derkachov:1997ch} that the contribution to an anomalous dimension
 coming from diagrams with self-energy insertions and vertex
corrections to a LO diagram  can be obtained by evaluation the LO diagram with dressed propagators and vertices. The dressed
propagators have the form~\footnote{The expressions for dressed propagators  given in the Appendix of
ref.~\cite{Manashov:2016uam} correspond to a different normalization condition for the propagator of the $\sigma$ field, and the
expression for the fermion propagator contains a typo.}
\begin{align}
D_q(x)= -\frac{\widehat A\,\slashed{x}}{x^{2\Delta_q}}\,, &&  D_\sigma(x)=-\frac1n \frac{\widehat B}{x^{2\Delta_\sigma}}
\end{align}
where
\begin{align}
\Delta_q=\alpha+\eta/2\,, && \Delta_\sigma=1+\gamma_\sigma=1-\eta-2\chi\,,
\end{align}
and up to $O(1/n^2)$ terms
\begin{align}
\widehat A &= A(\mu)\cdot M^{-2\gamma_q} \left(1-\frac{\gamma_q}\mu
\right)\,,
\notag\\
\widehat B &= B(\mu)\cdot M^{-2\gamma_\sigma} \left(1-\gamma_\sigma \left(B_1-\frac1{\mu(\mu-1)}\right)
\right)\,.
\end{align}
The dressed vertex is shown in Fig.~\ref{fig:Vertex}.
\vskip -5mm \noindent Its form is fixed by the conformal invariance
\begin{align*}
V(x,y,z)=\frac{\widehat Z}{(z-y)^{2(b-\chi)}}
\frac{\slashed{x}-\slashed{z}}{(x-z)^{2(a+1+\chi)}}\frac{\slashed{y}-\slashed{y}}{(y-x)^{2(a+1+\chi)}},
\end{align*}
\begin{figure*}[t]
\begin{center}
  \includegraphics[width=\linewidth]{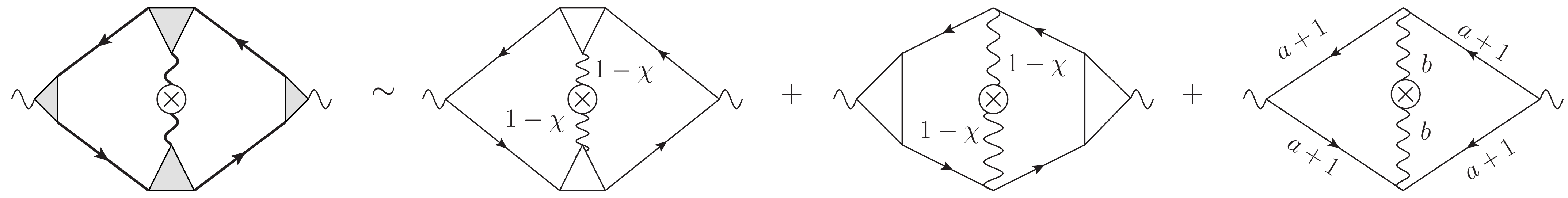}
\end{center}
\caption{ An example of the calculation of  the dressed diagram.}
  \label{fig:RHOMBUS}
\end{figure*}
where
\begin{align}
a =\Delta_q-1=\alpha+\eta/2\,,  && b=1-\eta\,,
\end{align}
$\chi=\kappa/2$ is defined in~\eqref{indices} and for the factor $\widehat Z$ we derived
\begin{align}
\widehat Z &= \pi^{-2\mu} M^{-2\chi}\cdot\chi\, a^2(1) (\mu-1)^3\left(1-\frac{\gamma_\sigma}{\mu-1}
\right)\,.
\end{align}
In order to obtain the contribution to an anomalous dimension from $1/n$ diagrams,\footnote{The diagrams have to satisfy the
condition
$N_V=N_q=2N_\sigma$, where $N_V$ is the number of the $\sigma\bar q q$ vertices, $N_a$ and $N_\sigma$ are the numbers of
fermion and $\sigma$-propagators, respectively.} see Fig.~\ref{fig:g-LO}, with all possible SE insertions and vertex
corrections (SEV) one has to replace all propagators and vertices in
the $1/n$ diagrams by the dressed ones and introduce a
regulator
$\Delta$ in one of the lines, e.g. to shift the index of the fermion propagator, $\Delta_q \to \Delta_q-\Delta$. The
corresponding diagrams have only a superficial divergence. Then the contribution to the anomalous dimension
reads~\cite{Derkachov:1997ch}
\begin{align}
\gamma_1+\gamma_2^{(SEV)}=-2\sum_{i} r_i,
\end{align}
where $r_i$ is the residue at the $\Delta$ pole of the dressed diagram, $D_i(\Delta)=r_i/\Delta +O(\Delta^0)$ and the sum goes
over all such diagrams (in the case under consideration $i=1,2$). Proceeding in this way one reduces the number diagrams to be
calculated. Moreover, individual vertex correction diagrams appear to be more complicated: all of them have a divergent subgraph.
The presence of the regulator
$\Delta$  breaks the uniqueness condition for the basic scalar-fermion vertices that rather complicates a calculation.
On contrary, all vertices in a dressed diagram obey the uniqueness condition and a diagram can be transformed with the help of
the star-triangle relation. In order to further simplify the calculation one can  dress only part of the propagators and
vertices. Indeed, the residues $r_i$ depend on two parameters, $\eta$ and $\chi$, which we will, for a while, consider as
independent parameters. Since $r_i$ is needed up to $1/n^2$ terms one can write
\begin{align}
r(\eta,\chi)= r(0)+\eta r_\eta(0)+\chi r_\chi(0)+\ldots\,,
\end{align}
where $r_\eta=\partial_\eta r$, etc. It means that up to $1/n^2$ terms the calculation of the diagram $D(\eta,\chi)$ is
equivalent to the calculation of two simpler diagrams: $D(\eta,\chi=0)$ and $D(\eta=0,\chi)$. Note, that the parameter $\chi$
plays the role of a regulator in the vertex $V$. It can be checked that in the limit
$\chi\mapsto 0$ the vertex reduces to the point-like basic  vertex with unit coefficient.
Thus the diagram $D(\chi=0,\eta)$ coincides with the $1/n$ diagram with modified propagators and point-like vertices.

Moreover, taking into account that   contributions from  dressed vertices and propagators to an anomalous dimension are
additive (at the order
$1/n^2$) one can, in order to simplify a calculation, modify only part of the vertices and propagators.
A more general discussion of this technique can be found in~\cite{Manashov:2017xtt} while here we illustrate it on the example of the
calculation of the right diagram in Fig.~\ref{fig:g-LO}.

The corresponding dressed diagram is shown on the l.h.s in Fig.~\ref{fig:RHOMBUS} -- the fat lines stand for the dressed
propagators and the grey triangles for the dressed vertices. One needs to find the corresponding residue, $r(\chi,\eta)$.
 Since we  restrict ourselves to the $1/n^2$ accuracy the corresponding result can be extracted from another three diagrams,
see Fig.~~\ref{fig:RHOMBUS}. Namely,
\begin{align}
r(\chi,\eta)=r_1(\chi)+r_2(\chi)+r_3(\eta)-2r(0,0)\,.
\end{align}
The last diagram which is $D(0,\eta)$ depends only on $\eta$. The sum of the previous two diagrams gives $D(\chi,0)+D(0,0)$ up to
$O(1/n^3)$ terms. The white triangles stand for the dressed vertex $V|_{\eta=0}$ and lines without indices -- for the
bare  propagators.

Let us start the analysis from the last diagram, which we denote by $D_3(a,b)$. Since the indices $a=\mu-1+\eta/2$, $b=1-\eta$ obey the
uniqueness condition
$2a+b=2\mu-1$ the diagram can be easily calculated.
For the corresponding residue we get
\begin{align}
r_3(\eta) &= N_\eta^2 R(\eta)\,,
\end{align}
where
\begin{align*}
N_\eta=\widehat A^2  \widehat B \Big|_{\chi=0}=A^2
 B
 \left(1+\eta\left(B_1-\frac1{\mu-1}\right)\right),
\end{align*}
and
\begin{align}
 R(\eta) &=
\frac{\pi^{4\mu}}{\Gamma(\mu+1)} 
\left(\frac{\Gamma(a')\Gamma(b')}{\Gamma(a+1)\Gamma(b)}\right)^4
\frac{\Gamma(2b+1-\mu)}{\Gamma(2b'-1)}
\notag \\
&\quad
\times\Biggl[1-\frac{2b+1-\mu}{b}\left(1-\frac{2a}{b}\frac{a'}{b'-1}\right)\Biggr]
\end{align}
with $x'\equiv \mu-x$.

With the help of the star-triangle relation  the second (four-loop) diagram  can be transformed (up to a prefactor) into the
diagram
$D_3(a_\chi,b_\chi)$,
where
$a_\chi=\alpha+\chi$ and $b_\chi=1-2\chi$, that gives for the residue $r_2$
\begin{align}
r_{2} &=
\frac{\pi^{4\mu}}{\Gamma^4(\mu)}
\frac{ A^2(\mu) B(\mu)\, V_\chi R(2\chi)}{\chi^2(\mu-1+\chi)^2} H(\chi)/H(2\chi)\,,
\end{align}
where $H(\chi)=a^2(1-\chi)a(2\mu-3+2\chi)$ and
\begin{align}
V_\chi =\widehat A^2 \widehat Z^2 \widehat B \Big|_{\eta=0} &=
\frac{\chi^2 M^{-4\chi}}4 \left(\frac{a(1)}{\pi^\mu}\right)^6 \alpha^{8} B(\mu)
\notag\\
&\quad \times
\left(1+2\chi \left(B_1+\frac{\mu+\alpha}{\mu\alpha}\right)\right)\,.
\end{align}
The calculation of the first diagram follows the same lines. Using the star-triangle relation one can show that
$$
D_1\sim D_3(\alpha+\chi,1-\chi)= D_3(\alpha+\chi,1-2\chi) \frac{D_3(\alpha,1+\chi)}{D_3(\alpha,1)}.
$$
It results in the following expression for $r_1$
\begin{align}
r_1=\frac{\pi^{4\mu}}{\Gamma^4(\mu)}
\frac{ A^2(\mu) B(\mu)\, V_\chi R(2\chi)}{\chi^2(\mu-1+\chi)^2} E(\chi)\,,
\end{align}
where the factor $E(\chi)={D_3(\alpha,1+\chi)}/{D_3(\alpha,1)}+O(1/n^2)$ has the form
\begin{align}
E(\chi)&= \Big(1+2\chi B_3\Big)
\biggl\{\frac{\mu^2-4\mu+2}{\mu-2}+\chi(\mu-1)\times
\notag\\
&\quad\left[\frac{3\mu}{2\mu-3} C_1-\frac{2(\mu^2-4\mu+6)}{(\mu-2)^2}\right]\biggr\}\,.
\end{align}
The calculation of the first diagram in Fig.~\ref{fig:g-LO} is much simpler. The final answer for both diagrams with
self-energy insertions and vertex corrections are given in~\ref{app:part_results}.

\section{Results for individual diagrams}\label{app:part_results}
In this appendix we collect the  expressions for all Feynman diagrams needed for our analysis. We give the divergent part of
each diagram after subtraction of subdivergences, symmetry factors being already included.

 {\allowdisplaybreaks
For the diagrams in Fig. \ref{fig:Dgamma3} we obtain
\begin{align}
D_1 &= 
-\frac{u^3}{\Delta} \eta_1^2 
{\mu^2(2\mu-3)}/\alpha\,,\notag
\\
D_2 &= 
-\frac{3u^3}{\Delta}\eta_1^2 \mu^2 C_1\,, \notag
\\
D_3  &= 
-\frac{u^3}{\Delta}\eta_1^2 {\mu^2} C_1/\alpha\,,
\\
D_{4+5} &= 
\frac{3u^4}2 \frac{\eta_1^2\mu^2(2\mu-1)}{\alpha}\left\{-\frac1{\Delta^2}+\frac2\Delta\right\}\,,
\notag
\\
D_{6+7} &= 
\frac{3u^4}4\frac{\eta_1^2\mu^2(2\mu-1)}{\alpha^2} \left\{-\frac1{\Delta^2}+\frac1\Delta
\left[\frac2{\alpha}+3\alpha \,C_1\right]\right\}\,.\notag
\end{align}
The diagrams in Fig.~\ref{fig:OC} give the following contributions
\begin{align*}
D_\text{OC} =&
-\frac{u^4}{32}\widehat{\gamma}_1^{\phantom{2}2}
\,\Bigg\{
\frac1{\Delta^2}+\frac{1}{\Delta}
\Big[2 B_3 -\frac{3 \alpha  \left(\alpha ^2-1\right) C_1}{4 \alpha^2-1}
\notag \\
&+ \frac{\alpha ^5-4 \alpha ^4-5 \alpha ^3+10 \alpha ^2+8 \alpha +2}
{\alpha  (2 \alpha +1)
 \left(\alpha ^2-1\right)}
\Big]
\Bigg\}\,.
\end{align*}
For the single box diagrams we obtain:
\begin{align*}
SB_1  &= \frac{u^3\eta_1^2}{12\Delta}\frac{\mu \alpha(2\alpha-1)}{\alpha-1}\,,
\notag\\
SB_2 &= -\frac{u^3\eta_1^2}{3\Delta} \frac{\mu (2\alpha-1)}{\alpha(\alpha-1)}
\left[-1+(\alpha-1)^2-\frac1{\alpha^2}\right]\,,
\notag\\
SB_3 &= \frac{u^3\eta_1^2}{2\Delta}\frac{\mu}{\alpha}\Big[
(\alpha^2 - \alpha + 1) C_1
\notag\\
&\quad +\frac23\frac{(2 \alpha -1) \left(\alpha ^3-\alpha ^2-\alpha -1\right)}{\alpha ^2}
\Big]\,.
\end{align*}
Finally, our results for double box diagrams (Fig. \ref{fig:DoubleBoxes}) are
\begin{align*}
DB_1 =& \frac{u^4}\Delta \frac{\mu(2\alpha-1)}{\alpha(\alpha-1)}
\eta_1\bigg\{
\frac{\left(2 \alpha ^4-4 \alpha ^3+\alpha ^2-2 \alpha +1\right)}
{\alpha +1}\notag
\\
&+\eta_1\bigg[
\frac{\left(2 \alpha ^3-7 \alpha ^2+4 \alpha -1\right) B_3}{4\alpha}
-\alpha ^2-\frac{1}{2 \alpha ^2}\notag \\
&+\frac{11 \alpha }{4}-\frac{1}{2 (\alpha -1)}
-\frac{1}{4 (2\alpha -1)}+\frac{1}{\alpha }-2
\bigg]\bigg\} \,,
\\[2mm]
DB_2 =&\frac{u^4 }\Delta \frac{\mu (2\alpha-1)}{2\alpha}\eta_1^2
\bigg\{
\frac{2 \alpha ^4-\alpha ^3-\alpha ^2+3 \alpha -1}{\alpha^2(2\alpha-1)
}
\\
&\quad +\frac{1
}
{4} \mathcal{B}
+\frac{7
}{4
} C_1
-\frac{
2 \alpha ^3-\alpha ^2-3 \alpha +1
}{2\alpha(\alpha-1)} B_3 
\bigg\}\,, 
\\[2mm]
DB_3  =&
-\frac{2u^4}\Delta \mu \alpha 
(2\alpha-1)\eta_1\bigg\{
-\frac{
2 \alpha ^2-2 \alpha +1}
{(\alpha -1) (\alpha +1)} \notag
\\
&+ \frac{\eta_1}{4\alpha^2}\biggl[
\frac{1
}{2
} \Big(\mathcal{B}+C_1\Big)
+\frac{
 3 \alpha -1}{
 \alpha(\alpha -1) }  B_3
\notag \\
&
+\frac{4 \alpha ^5-8 \alpha ^4+\alpha ^3+12 \alpha ^2-10 \alpha +2}{\alpha (\alpha -1)^2 (2\alpha-1)
}
\biggr]
\bigg\}\,,
\\[2mm]
DB_4=& \frac{u^4}{2\Delta}\mu \alpha 
(2\alpha-1)\eta_1\bigg\{
\frac{2 \alpha  
\left(2 \alpha ^2-4 \alpha +3\right)}{(\alpha -2) (\alpha -1) (\alpha +1)}
\notag \\
&+\frac{\eta_1}{\alpha^2}\biggl[
\frac{1
}{4 
} \Big(\mathcal{B}+C_1\Big)
-\frac{2  (\alpha-1) 
}{\alpha -2 
} B_3
\notag \\
&-\frac{
2 \alpha ^4-9 \alpha ^3+14 \alpha ^2-10 \alpha +2
}{2(\alpha -2) (\alpha -1)^2 
}
\biggr]
\bigg\}\,.
\end{align*}
Here  $\mathcal{B}=B_3^2-C_{3-\mu}$.
}

The contribution to the anomalous dimension from a diagram $\mathcal{D}_a$ reads
\begin{align}
\gamma_{\mathcal{D}_a}= -\partial_u \mathcal{D}_a^{(1)}|_{u=1}\,,
\end{align}
where $\mathcal{D}_a^{(1)}$ is the simple pole residue, $\mathcal{D}_a=\mathcal{D}_a^{(1)}/\Delta+\ldots$.

Finally we give the contributions to the anomalous dimensions from the self-energy and vertex correction diegrams
\begin{align*}
\gamma_{{SEV}_1} &= 
\frac12\eta_1^2
\left(\frac{34\alpha^4+5\alpha^3-55\alpha^2+6\alpha+4}{\alpha(\alpha-1)}\right)\,,
\notag\\
\gamma_{{SEV}_2} &= 
\eta_1^2\Biggl\{-\frac{3(\alpha-1)(\alpha+1)^2}{\alpha} C_1
\notag\\
&\quad
-\frac{2(10\alpha^6-7\alpha^5-27\alpha^4+12\alpha^3+6\alpha^2-\alpha+1)}{\alpha^3(\alpha-1)}\Biggr\},
\end{align*}
where $\gamma_{\text{SEV}_1}$, $\gamma_{\text{SEV}_2}$ correspond to the left and  right diagrams in Fig.~\ref{fig:g-LO},
respectively.

\end{document}